# Superradiance of Spin Defects in Silicon Carbide for Maser Applications

Andreas Gottscholl[1], Maximilian Wagenhöfer[1], Manuel Klimmer[1], Selina Scherbel[1], Christian Kasper[1], Valentin Baianov[1], Georgy V. Astakhov[2], Vladimir Dyakonov[1]*, Andreas Sperlich[1]*

[1]Experimental Physics 6 and Würzburg-Dresden Cluster of Excellence, Julius Maximilian University of Würzburg, Würzburg, Germany

[2]Helmholtz-Zentrum Dresden-Rossendorf, Institute of Ion Beam Physics and Materials Research, 01328 Dresden, Germany

*Corresponding Authors: dyakonov@physik.uni-wuerzburg.de, sperlich@physik.uni-wuerzburg.de

Keywords: Stimulated Emission, Maser, Population Inversion, Silicon Vacancy, Spin Polarization

## Abstract

Masers as telecommunication amplifiers have been known for decades, yet their application is strongly limited due to extreme operating conditions requiring vacuum techniques and cryogenic temperatures. Recently, a new generation of masers has been invented based on optically pumped spin states in pentacene and diamond. In this study, we pave the way for masers based on spin $S = 3/2$ silicon vacancy ($V_{Si}$) defects in silicon carbide (SiC) to overcome the microwave generation threshold and discuss the advantages of this highly developed spin hosting material. To achieve population inversion, we optically pump the $V_{Si}$ into their $m_S = \pm 1/2$ spin sub-states and additionally tune the Zeeman energy splitting by applying an external magnetic field. In this way, the prerequisites for stimulated emission by means of resonant microwaves in the 10 GHz range are fulfilled. On the way to realising a maser, we were able to systematically solve a series of subtasks that improved the underlying relevant physical parameters of the SiC samples. Among others, we investigated the pump efficiency as a function of the optical excitation wavelength and the angle between the magnetic field and the defect symmetry axis in order to boost the population inversion factor, a key figure of merit for the targeted microwave oscillator. Furthermore, we developed a high-Q sapphire microwave resonator ($Q \approx 10^4 - 10^5$) with which we find superradiant stimulated microwave emission. In summary, SiC with optimized spin defect density and thus spin relaxation rates is well on its way of becoming a suitable maser gain material with wide-ranging applications.

## Introduction

Masers are established systems in deep-space communications and navigation due to their application as low-noise amplifiers and oscillators [Gordon et al. 1955]. However, their field of application is limited to these niche utilizations because of the required operation conditions such as low temperatures and high vacuum [Benmessai et al. 2008]. A new generation of solid-state room-temperature masers can offer a solution here [Arroo et al. 2021]. A few years ago, a maser at room-temperature was demonstrated for optically pumped pentacene-doped p-terphenyl [Oxborrow et al. 2012; Breeze et al. 2015; Breeze et al. 2017; Salvadori et al. 2017]. Its operating mode is restricted to a pulsed output, as the organic material, for instance, would be damaged at the required high pump power (maser threshold ≈230 W). This was followed by the demonstration of a diamond-based maser using NV-centres where continuous maser operation could be realized [Jin et al. 2015; Breeze et al. 2018]. Silicon carbide (SiC) is being considered as another potential material for masers, as it hosts optically addressable high-spin defects and can be manufactured on a large scale in various polytypes due to its industrial importance in high-power electronics. [Kraus et al. 2014]. If a SiC-based maser succeeds, this technologically mature material system can quickly find its way into everyday applications. Besides the technological advantages of SiC, it also has disadvantages such as shorter spin coherence times or peculiarities related to its spin quartet energy level structure in the case of the $V_{Si}$ defect, which lead to a less efficient pumping scheme [Wimbauer et al. 1997; Carter et al. 2015; Janzén et al. 2009]. To overcome these challenges, optimized spin pumping is essential which will be demonstrated in this work. We studied the influence of the excitation wavelength and the orientation of defects in the external magnetic field by means of the population inversion – the figure of merit determined via electron paramagnetic resonance (EPR). Finally, in combination with a high-Q resonator ($Q \approx 10^4 - 10^5$), we demonstrate a superradiant amplification of the incident microwaves, analogous to the superluminescence of laser diodes in the optical spectrum, which is a precursor to a functional coherent microwave emitter. Finally, we discuss the missing pieces of the puzzle to realize a SiC maser.

## Optical Spin Polarization of V2 in SiC

Silicon carbide possesses a large variety of different spin defects in its crystal lattice [Vainer et al. 1981; Baranov et al. 2005; Son et al. 2006; Koehl et al. 2011]. For the present work, the most relevant defect type is the negatively charged silicon vacancy ($V_{Si}^-$) at the cubic lattice site position in a 4H SiC lattice (in the following denoted as V2). It forms a spin quartet ($S = 3/2$) ground state with a zero-field splitting (ZFS) of $\approx 70$ MHz [Orlinski et al. 2003]. The spin system can be optically excited, and the excited state subsequently relaxes to the ground state (GS) in two ways: either radiatively by photoluminescence (PL) or non-radiatively by intersystem crossing (ISC) via a metastable state. Here, spin-selective transitions occur resulting in a spin polarization of the $\pm 1/2$ GS (Fig. 1A) [Kraus et al. 2014].

The prerequisite for the maser is – as for the laser – a population inversion of the states where the stimulated emission can occur. This can be achieved, when the pump rate $\omega$ of the excitation source exceeds the relaxation rate $\gamma_{eg}$ between the involved sublevels. The complete maser threshold can be described by Eq. (1) [Jin et al. 2015]:

$$Q \geq \frac{\omega + \gamma_{eg}}{\omega - \gamma_{eg}} \frac{\kappa_S \omega_c}{4Ng^2} \quad (1)$$

In addition to critical parameters such as the decay rate $\kappa_S$ of the spin collective mode (significantly limited by $T_2^*$) and the spin-photon-coupling $g$, a large number of spins $N$ and a resonator with a high Q-factor (with resonance frequency $\omega_c$) are of great importance (see [Jin et al. 2015] for more details). To accomplish the condition of population inversion, an external static magnetic field is applied which lifts the degeneracy of the $\pm 1/2$ and $\pm 3/2$ states due to the Zeeman effect. Upon optical pumping this results in a population inversion between the $-1/2$ and $-3/2$ states, which will be characterized in the following.

### Electron paramagnetic resonance

Fig. 1B illustrates the used EPR spectrometer, which is the basic tool to examine the spin population differences [Fischer et al. 2018; Breeze et al. 2018]. The SiC sample is placed in a cylindrical dielectric sapphire resonator inside a copper cavity ($Q \approx 17000$ at ambient temperatures) and can be optically pumped using an 808 nm laser. An external magnetic field is applied to tune the Zeeman splitting between the spin sublevels into the range corresponding to the resonance frequency of the resonator. A home-made microwave bridge is used to determine and to optimize the spin polarization of the spin defects. An Anritsu source generates microwaves with a frequency of ≈9.3 GHz that are then attenuated to a sufficiently low power to avoid saturating the transitions ($P_{MW} \approx$ -80 dBm = 10 pW). The microwaves are guided via a circulator to a loop antenna in the cavity, where they interact with the spin defects. The circulator directs the reflected microwaves to a low-noise amplifier before they are fed into a mixer with phase-adjusted local oscillator. A sinusoidal modulation of the external magnetic field results in a modulated mixer output that is detected by a lock-in amplifier. The resulting EPR signal is the first derivative of the sample's microwaves absorption spectrum (Fig. 1C). The V2 EPR spectrum consists of three features corresponding to the $\Delta m_s = \pm 1$ transitions illustrated in Fig. 1A. Upon optical pumping, we can observe two absorption peaks (blue and black) and one emission peak (red). The distinction between absorption and emission is based on the opposite sign of the EPR signals, as explained below. Each of the most pronounced EPR peaks ($B_-$, $B_0$, $B_+$) has multiple satellite peaks corresponding to the surrounding $^{29}$Si isotopes, which have a nuclear spin of $I = 1/2$ [Mizuochi et al. 2002]. The intensity of the central $B_0$ peak is not influenced by illumination but by temperature, resulting in a population difference of $\pm 1/2$ spin states according to Boltzmann distribution. The intensity of the $B_-$ peak (blue) is related to the spin polarization of the system, showing on the one hand the sensitivity of EPR for monitoring the population difference, but more importantly the optical pump efficiency. Noteworthy is that the high-field EPR feature $B_+$ (red) reveals the same intensity, but with an inverted phase – the signal first goes down and then up, i.e., it corresponds to microwave emission instead of absorption as in the case of the $B_-$ transition. The two insets in Fig. 1C illustrate this mirror-image behaviour not only for the two transitions but also for the hyperfine (HF) satellite peaks, which exhibit identical absorption and emission behaviour. Both satellite signals can be perfectly fitted using the derivative of three Lorentzians weighted with the natural abundance of $^{29}$Si (4.7%), as shown by the black dashed traces in the insets.

The figure of merit for characterizing the spin polarization is the population difference $\Delta p^\pm$ which can be determined for the absorptive transition at $B_-$ and the emissive transition at $B_+$, respectively.

$$\Delta p^\pm = \frac{I_{exc}^\pm}{I_{dark}^\pm} \Delta p_B \quad (2)$$

Here, the peak-to-peak amplitudes of the EPR spectra $I_{exc}^\pm$ of the optically excited sample are normalized to the amplitudes $I_{dark}^\pm$ of a dark measurement, which corresponds to the population difference $\Delta p_B$ due to Boltzmann statistics and serves as a scaling factor. To achieve population inversion, the optical pumping process must counteract thermodynamic equilibrium with the population difference described by Boltzmann statistics. Therefore, the population inversion $\Delta p^+$ of the emissive EPR transition at $B_+$ is reduced by $\Delta p_B$. We observe a symmetrical behaviour of the population differences $\Delta p^-$ and $\Delta p^+$ for different laser powers as depicted in the bottom inset of Fig. 1C. Both are symmetric around the Boltzmann population $\Delta p_B$ (black dashed line) with a theoretical maximum $|\Delta p^\pm| = 50\%$ for full polarisation to the $\pm 1/2$ states [Fischer et al. 2018]. As soon as the laser pumps the system sufficiently $|\Delta p^\pm| \gg \Delta p_B$, $\Delta p_B$ can be neglected and we observe $\Delta p^- \approx |\Delta p^+|$. An identical intensity for both transitions is therefore an indicator for a very efficient optical pumping. In the following sections, we will analyse the spin polarization for various external influencing factors.

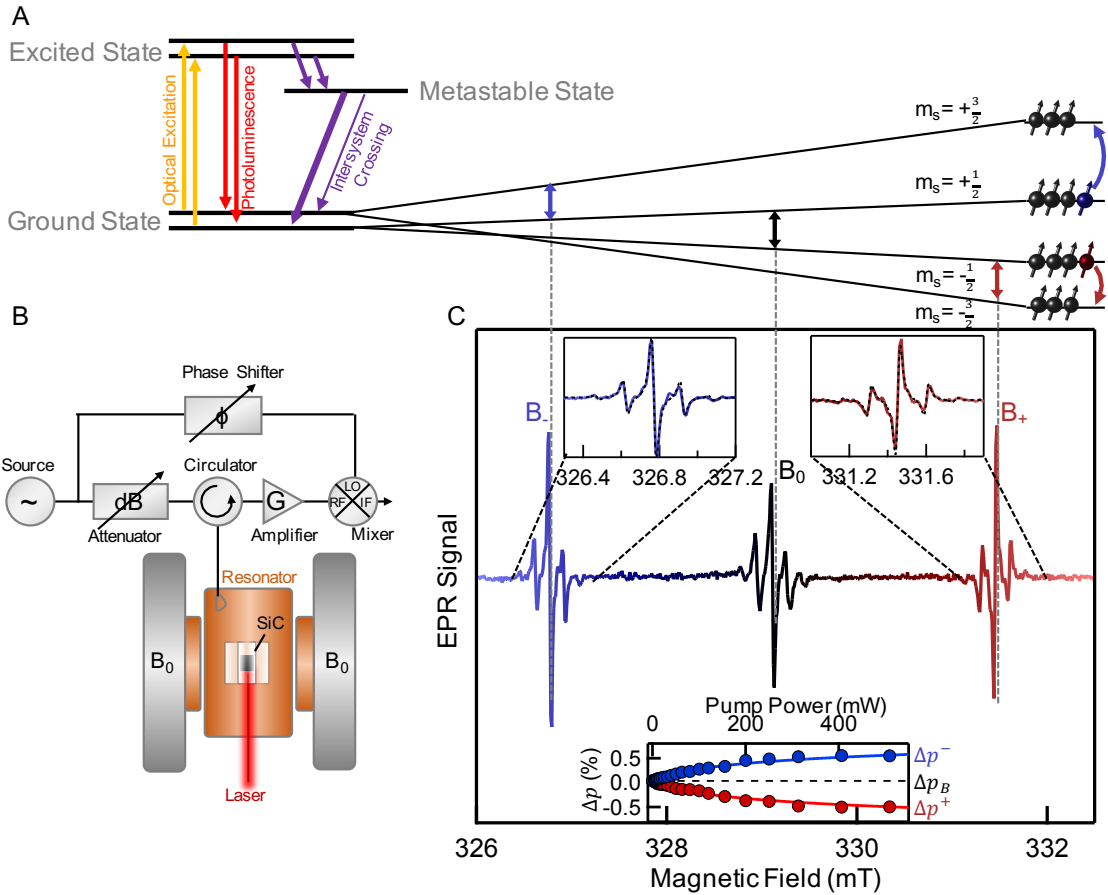

**Figure 1:** Optical pumping and stimulated emission of the V2 spin defect in 4H SiC: **(A)** Schematic of the optical excitation cycle leading to a preferential population of the ±1/2 GS sublevels. The application of an external magnetic field leads to a splitting of the energy levels due to the Zeeman effect and results in a population inversion between the −1/2 and the −3/2 states. **(B)** EPR setup for detecting microwave emission from a laser-pumped SiC sample placed in a resonator. Note that the resonant microwaves also serve to induce stimulated emission. **(C)** EPR spectrum of V2 in 4H SiC under optical excitation at room-temperature. The low ($B_-$) and high ($B_+$) field EPR transitions show different phase, indicating population inversion under optical excitation, while the signal at $B_0$ is insensitive to light. The bottom inset displays the pump power dependency of the population differences at $B_-$ (blue) and $B_+$ (red) in relation to the Boltzmann population (black dashed line).

**Wavelength-dependent optical pumping**

One way of optimizing the spin polarization of $V_{Si}$ is to vary the wavelength of optical excitation. Obviously, it is more efficient to resonantly excite the spin centre directly at the zero-phonon line (ZPL) absorption instead of in the broad phonon side bands. However, this effect is strongly temperature-dependent. In Fig. 2A, PL spectra of 4H SiC are shown for several temperatures in the range of 10 – 300 K. At cryogenic temperatures, a zoo of sharp ZPLs appears. They can be attributed to different optically active defects, such as Frenkel and Schottky defects, divacancies and silicon vacancies located at different lattice sites [Vainer et al. 1981; Baranov et al. 2005; Son et al. 2006; Koehl et al. 2011]. For our EPR studies we focus on the sharp line located at 917 nm which is attributed to the V2 defect [Wagner & Bechstedt 2000]. However, as can be clearly seen in Fig. 2A, the ZPL is only detectable at low temperatures and disappears into the phonon background at temperatures between 50 and 100 K.

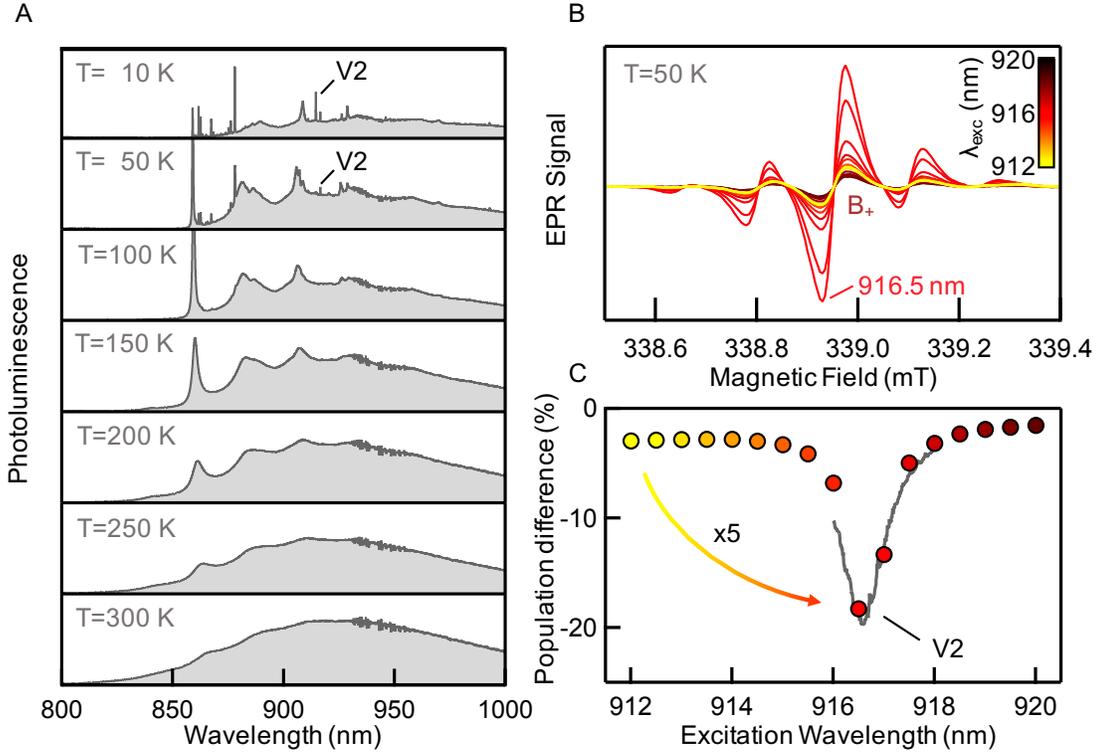

**Figure 2:** Resonant optical excitation of V2 in 4H SiC: **(A)**. Photoluminescence spectra under 633 nm excitation at different temperatures (10 – 300 K). At low temperatures V2 emits a sharp ZPL at 917 nm. **(B)** EPR spectra taken at different excitation wavelengths, as indicated in the legend. **(C)** Population difference as a function of excitation wavelengths at $T = 50$ K. The values (circles) are calculated by Eq. (2) using single EPR measurements performed with the respective excitation wavelength (same colour code as in B). The grey trace is the normalized EPR intensity during a sweep of the excitation wavelength. Both measurements show efficient optical pumping for resonant excitation at the ZPL wavelength.

In Fig. 2B, we show the dependence of the emissive $B_+$ transition EPR spectrum on different excitation wavelengths in steps of 0.5 nm at $T = 50$ K and laser power of 10 mW. The highest spin polarization is achieved at 916.5 nm which coincides with the literature value of the ZPL of ca. 917 nm [Wagner & Bechstedt 2000]. Fig. 2C shows the resulting population difference, which was calculated by Eq. (2) [Fischer et al. 2018]. Each circle represents one EPR measurement with colours corresponding to the spectra in Fig. 2B. At a wavelength matching the ZPL, the population inversion increases by a factor of 5. To rule out the possibility of additional features appearing between the individual measurements, we performed EPR measurements at a fixed magnetic field ($B_+$) while the wavelength of the laser excitation was continuously adjusted (grey trace). The profile coincides perfectly with the single measurements and does not show any additional features. Thus, we conclude that at low temperatures, resonant optical pumping at the ZPL is particularly efficient.

At higher temperatures, the ZPL of V2 vanishes and the effect of resonant excitation can be neglected, suggesting using photons with higher energy than the ZPL (Stokes excitation). Anti-Stokes excitation, e.g., above 1000 nm exploiting phonons is also conceivable, but its advantages, such as high ODMR contrast, are expected only well above room temperatures, making such excitation more suitable for realizing a maser at $T > 300$ K [Wang et al. 2021]. To take advantage of a prolonged spin-lattice relaxation time and thus high spin polarization, we will continue to search for an optimal set of physical parameters at temperatures below room temperature and thus using Stokes excitation (808 nm).

**Influence of SiC crystal orientation in the magnetic field**

Another parameter that strongly affects the spin polarization is the polar angle $\theta$ of the SiC crystal axis $\vec{c}$ and thus the spin defects with respect to the external magnetic field $\vec{B}$. The angular dependence of the resonance magnetic fields of EPR transitions for high-spin states including those in SiC has been known for years [Weiland & Bolton 2008; Baranov et al. 2001].

$$\Delta B = \frac{1}{g\mu_B} \left\| \underline{D} \cdot \frac{\vec{B}}{|\vec{B}|} \right\| = \frac{2}{g\mu_B} D(3\cos^2\theta - 1) \quad (3)$$

Here, $g$ is the g-factor, $\mu_B$ is the Bohr magneton, $\underline{D}$ is the ZFS tensor with its diagonal elements $\left(\frac{1}{3}D, \frac{1}{3}D, -\frac{2}{3}D\right)$ and $D$ is the corresponding scalar ZFS parameter. EPR spectra for different angles $\theta$ are displayed in Fig. 3A. The distance $\Delta B$ between the two outer peaks $B_-$ and $B_+$ reveals the standard dependence according to $\theta$. Here we show that the spin polarization shows the same anisotropy. The experimentally determined splitting $\Delta B$ is displayed in Fig. 3B (dark grey circles) together with the expected trajectory (solid trace) according to Eq. (3). A special case is given at $\theta \approx 54.7°$, where the splitting is zero, i.e., $B_-$ and $B_+$ transitions coincide at $B_0$. Interestingly, extracting the population inversion from these angular-dependent EPR measurements shows the same behaviour (see Fig. 3C). The dark red circles represent the measured data. Although full rotation was not possible due to the coupled optical fibre, for symmetry reasons we mirrored the data (light red circles) to complete the picture of the angle-dependent spin polarization. Remarkably, an identical behaviour of $\Delta B$ and spin polarization can be observed. This apparently intuitive finding is not at all obvious, and according to our knowledge, has not been demonstrated so far. This appears to be a general effect which can also be observed in other optically polarized high-spin material systems. For instance, we recently observed this effect qualitatively for the $V_B^-$ spin defect ($S = 1$) in hexagonal boron nitride (hBN) at low temperatures [Gottscholl et al. 2020]. With the strong data shown here, we can now analyse the anisotropy effect with much higher confidence. Plotting the data from Fig. 3B against Fig. 3C reveals a direct proportionality between $\Delta p^+$ and the angular-dependent splitting $\Delta B$ up to its maximum value of 2× *ZFS* for $\vec{B} \parallel \vec{c}$ as shown in Fig. 3D. To understand this behaviour, we will now discuss possible factors.

One possible candidate would be the magnetic field-dependent Boltzmann distribution given by $\exp(-g\mu_B B/k_B T)$ with the Boltzmann constant $k_B$ and temperature $T$. The corresponding population difference $\Delta p$ between two spin sublevels is larger at stronger magnetic fields and would therefore vary with the $\Delta B$ splitting. However, we observe a mirror-image behaviour for the absorptive transition $B_-$. Furthermore, the maximum $\Delta B$ splitting is small compared to the externally applied fields ($\Delta B_{max}/B_0 \approx 1\%$). Therefore, the influence of the Boltzmann statistics can be neglected. Another possible explanation might be a difference in the effective illumination intensity of the sample including the laser polarisation. In conventional EPR spectrometers, such as the one we use, excitation occurs through an optical window, resulting in a beam path that is perpendicular to the external magnetic field and the sample rotation axis. The effective illumination intensity on the sample is then described by $\sin(\theta)$, which contradicts the relationship $3\cos^2\theta - 1$. To minimize the influence of angle-dependent illumination, we replaced the excitation through the window with an excitation from the top by using a fibre-coupled sample holder made of a quartz rod. The use of an optical fibre also eliminates any polarisation of the used laser. Therefore, we assume identical excitations for all measurements shown in Fig. 3.

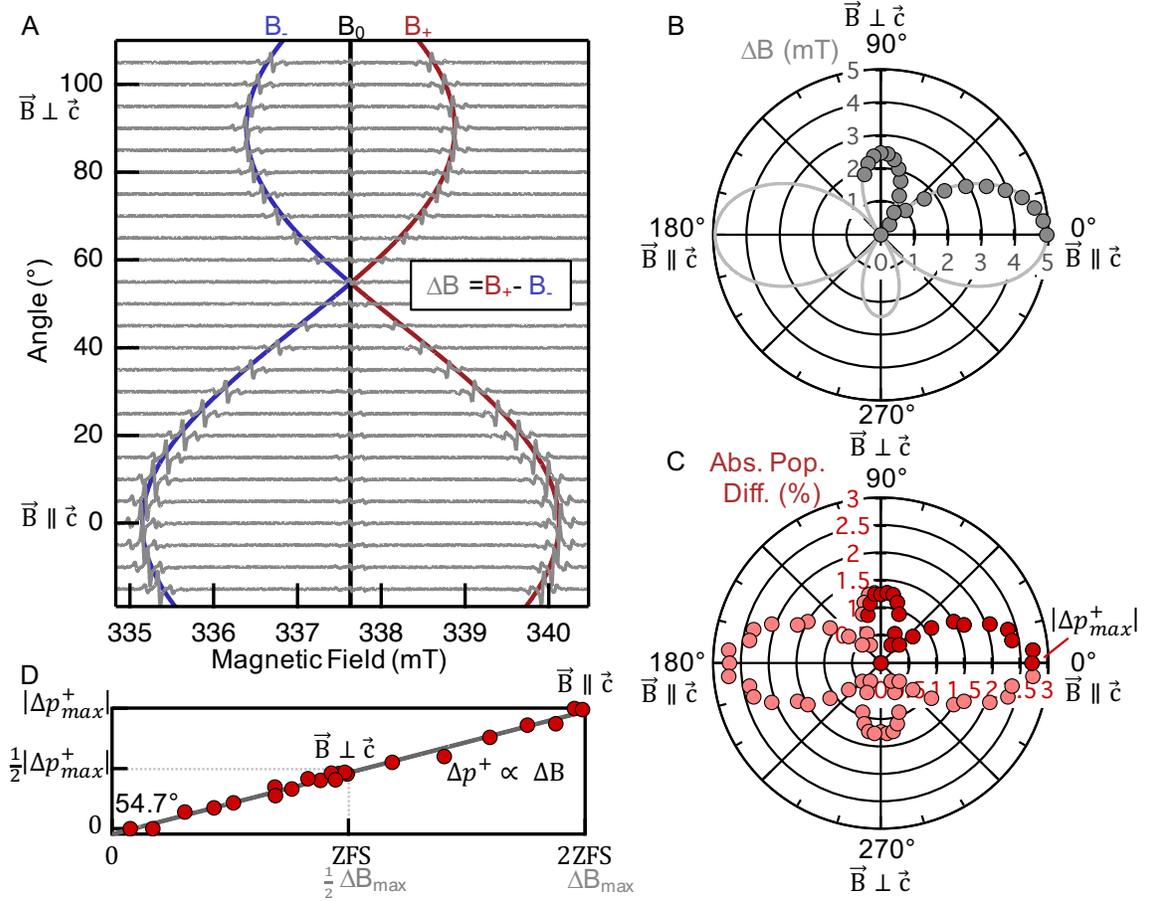

**Figure 3:** Anisotropy of V2 in 4H SiC: **(A)** Room temperature EPR spectra with fits (solid traces) by Eq. (3) for different angles between the external field $\vec{B}$ and the crystal $\vec{c}$-axis. The transitions coincide for an angle of 54.7°. **(B)** Angle-dependent splitting $\Delta B$ in a polar plot. **(C)** Experimentally determined population difference $|\Delta p^+|$ (dark red circles). Light red circles display mirrored experimental data assuming full rotational symmetry. **(D)** Plotting the absolute population difference $|\Delta p^+|$ against the observed splitting $\Delta B$ in units of the ZFS results in a linear relationship with slope $|\Delta p^+_{max}| / (2 \cdot ZFS)$.

Previous works have shown that the spin interactions of NV$^-$ defects in diamond are also anisotropic and that the maximum polarization is achieved when the defect axis $\vec{c}$ is aligned parallel with an applied magnetic field $\vec{B}$ [Epstein et al. 2005; Tetienne et al. 2012]. More recently, the polarization of the NV$^-$ Zeeman sublevels as a function of magnetic field orientation relative to $\vec{c}$ has also been analysed theoretically [Drake et al. 2016]. An explanation provided is based on the assumption of constant polarization in the spin defect coordinate system (frame), where the spin Hamiltonian is well-defined, while the experimentally accessible population difference is considered in the laboratory frame, which undergoes rotation. Depending on the orientation of the defect in the external magnetic field, the defect frame and the laboratory frame differ by an angle $\theta$ corresponding to a Wigner rotation of the spin Hamiltonian with $ZFS = 2D$ for $S = 3/2$. Just as the ZFS tensor is projected onto the external differently oriented magnetic field, resulting in a splitting $\Delta B$ (Eq. (3)), the population difference can also be considered as a tensor and treated in the same way. Thus, the population difference tensor $\underline{P}$ and the ZFS tensor $\underline{D}$ are connected by:

$$\underline{P} = \frac{\Delta p_{max}}{2 \cdot ZFS} \underline{D} \quad (4)$$

With $\Delta p_{max}$ being dependent on the experimental conditions and the splitting $\Delta B$ being scaled in units of the ZFS. However, Eq. (4) is only a good approximation for when the applied magnetic field is considerably larger than the ZFS. Consequently, combining Eq. (3) and Eq. (4), the experimentally achievable population difference is $\Delta p = \frac{1}{2} \Delta p_{max}(3\cos^2\theta - 1)$. This leads to optimal spin polarization for $\theta = 0$ $(\vec{B} \parallel \vec{c})$. While the effect of the excitation wavelength is only important for low temperature application, the effect of crystal orientation seems to be a temperature-independent property and becomes also relevant for room temperature maser operation.

**Combination of optimized optical pumping with a high Q-factor resonator**

To enhance the interaction of microwave photons with the spin system in a resonator, a higher Q-factor is required. Thereby the Q-factor describes the dampening or energy losses and therefore can only be increased by reducing the losses of the standing waves in the resonator. This can be accomplished by reducing coupling losses using an undercoupled resonator and, additionally, by cooling the system to reduce the ohmic losses in the copper walls. At $T = 110$ K we thus reach a Q-factor of $40000 - 110000$ by varying the depth of the coupling loop in the resonator. Another practical side effect of low temperatures is the prolonged spin lattice relaxation time $T_1$ leading to a lower required optical pump power [Simin et al. 2017].

To investigate this in more detail, we examine an EPR spectrum for a Q-factor of 84000 as shown in Fig. 4A. We choose the optimal sample orientation from Fig. 3 $(\vec{B} \parallel \vec{c})$ to obtain the highest achievable population inversion. For excitation we use the standard Stokes excitation (808 nm) since the effect of resonant excitation is not present above 100 K as the ZPL vanishes between 50 K and 100 K (Fig. 2A). In Fig. 4B, we show the influence of the Q-factor on the emissive EPR transition $B_+$. A massive signal increase (factor of 2.5 for the central peak of $B_+$) is observable for higher Q-factors. It seems that with such high Q-factors and the optimized crystal orientation, we are approaching spin polarization near the maser threshold as discussed below.

A remarkable result is the tremendous population asymmetry between the absorbing $B_-$ and emitting $B_+$ transition ($|\Delta p^+| \approx 7 \cdot \Delta p^-$). This asymmetry in $\Delta p^\pm$ cannot be explained by the population inversion, as this can only result in the emission and absorption intensities being approximately equal as shown in the inset of Fig. 1C with $|\Delta p^+| \approx \Delta p^-$. Therefore, we assign the immense stimulated emission as a sign of superradiance (comparable to superluminescence of lasers) [Gross & Haroche 1982].

In order to emphasize that we are just surpassing the onset of superradiance, we examine the hyperfine structure of the zoomed-in region around $B_+$ as shown in Fig. 4C. We will use the different natural abundances of silicon isotopes to show that the onset is clearly surpassed for V2 centres with just spin-less nuclei in their vicinity, but just barely for the lower number $N$ of V2 interacting with spin-bearing Si isotopes. As pointed out earlier, the EPR transitions $B_+$ or $B_-$ are not single lines but have two equally separated satellites that emerge due to HF interactions of the electron spin of V2 centres with one $^{29}$Si, $I = 1/2$ nuclear spin each. The central peak, however, is due to V2 centres that have only spin-less isotopes in their vicinity ($^{28}$Si, $^{30}$Si with $I = 0$). To remind, the three major stable silicon isotopes, $^{28}$Si, $^{29}$Si and $^{30}$Si have abundances of 92.2%, 4.7%, and 3.1%, respectively [Meija et al. 2016]. Consequently, the probability of having only spin-less isotopes as the 12 silicon neighbours of a specific V2 centre is $(92.2\% + 3.1\%)^{12} = 56.1\%$. The probability of having just one out of 12 being $^{29}$Si is $12 \cdot 4.7\% \cdot (100\% - 4.7\%)^{11} = 33.2\%$, while the remaining 10.7% of V2 centres have multiple $^{29}$Si neighbours. The additional satellite features of the latter are weaker and do not contribute significantly to the observed spectrum. The influence of spin-bearing $^{13}$C isotopes ($I = 1/2$) with abundance of just 1.1% is also not taken into account. The ratio of 56.1% / 33.2% is therefore the basis of the simulated EPR spectra in Fig. 1C that is also added as the dashed grey trace in Fig. 4C. The fixed isotope abundances thus lead to a predictable intensity ratio of the central peak in comparison to the satellites. However, the central emission transition is much more dominant than expected (approximately factor of 4). This discrepancy between the ratio of signal amplitudes ($I_+/I_{HF}$) and the isotope ratio ($N_+/N_{HF}$) is in line with the onset of superradiance which depends superlinearly on the number of involved spins squared [Gross & Haroche 1982; Angerer et al. 2018]. Here, in this sample alone, we have two separate ensembles with different number $N$ of involved V2 spin centres that we can distinguish with a single measurement. This distinction within the same sample is possible due to the different nuclear isotope types providing different EPR signals, which were quantified individually. Comparing the intensities of the central emissive peak $I_+$ with the intensity of one of the HF peaks $I_{HF}$ we can write:

$$\frac{I_+}{I_{HF}} = \left(\frac{N_+}{N_{HF}}\right)^k \text{ with } k = 2 \text{ for superradiance.} \quad (5)$$

The ratio of the spin numbers is given by the isotope ratio $\frac{N_+}{N_{HF}} = \frac{56.1\%}{16.6\%} = 3.38$ where we use 33.2% / 2 = 16.6% for one of the two HF peaks. For the intensity ratio we observe $\frac{I_+}{I_{HF}} = 13.64$ leading to an exponent of $k = 2.1 \approx 2$ which satisfies the expected $N^2$ law of superradiance.

Furthermore, varying the applied laser power $P$ results in additional indications of superradiance as shown in Fig. 4D. The effect on the population difference can be described by:

$$\Delta p^{\pm}(P) = \Delta p^{\pm}_{max} \cdot \ln\left(\frac{P_0 + P}{P_0 + P_{\alpha}}\right), \quad (6)$$

with $\Delta p^{\pm}_{max}$ as the maximum population difference reduced by a logarithmic depth-dependent laser light absorption in the sample with the saturation power $P_0$ and the reduced laser power $P_{\alpha}$ due to light absorption in the sample (see Ref. [Fischer et al. 2018] for more details). The corresponding fit is shown in the top of Fig. 4D. For the absorptive transition $B_-$, the HF intensities of central and satellite peaks (blue circles and crosses) saturate at the same value of about ≈3%. Remarkably, the emissive transition $B_+$ and HF satellites reaches much larger values and no saturation. Interesting to note is that the population difference reaches about twice the value for the central peak (red circles) in comparison to the HF satellite peaks (red crosses), which is in line with superradiance for the larger number $N$ of V2 centres with spin-less isotopes.

Furthermore, for superradiance a narrowing of the EPR linewidth due to monochromatization of the microwave photons is also expected. An analysis of the extracted linewidth is shown in Fig. 4D (bottom). The observed EPR linewidth is typically in the range of 45 µT for the absorptive $B_-$ transition (blue symbols). In contrast, the linewidth of the emissive $B_+$ transition changes by a factor of 0.5 with rising pump power (red circles), while this effect is less pronounced for the HF peaks (red crosses).

All three observations, the superlinear behaviour of microwave photon emission intensity with the number of involved spin centres $N$, the increased population difference and the spectral narrowing of the microwave emission upon optical pumping are strong indicators of superradiance and, at the same time, evidence that we have almost reached the maser threshold with our SiC-based quantum system.

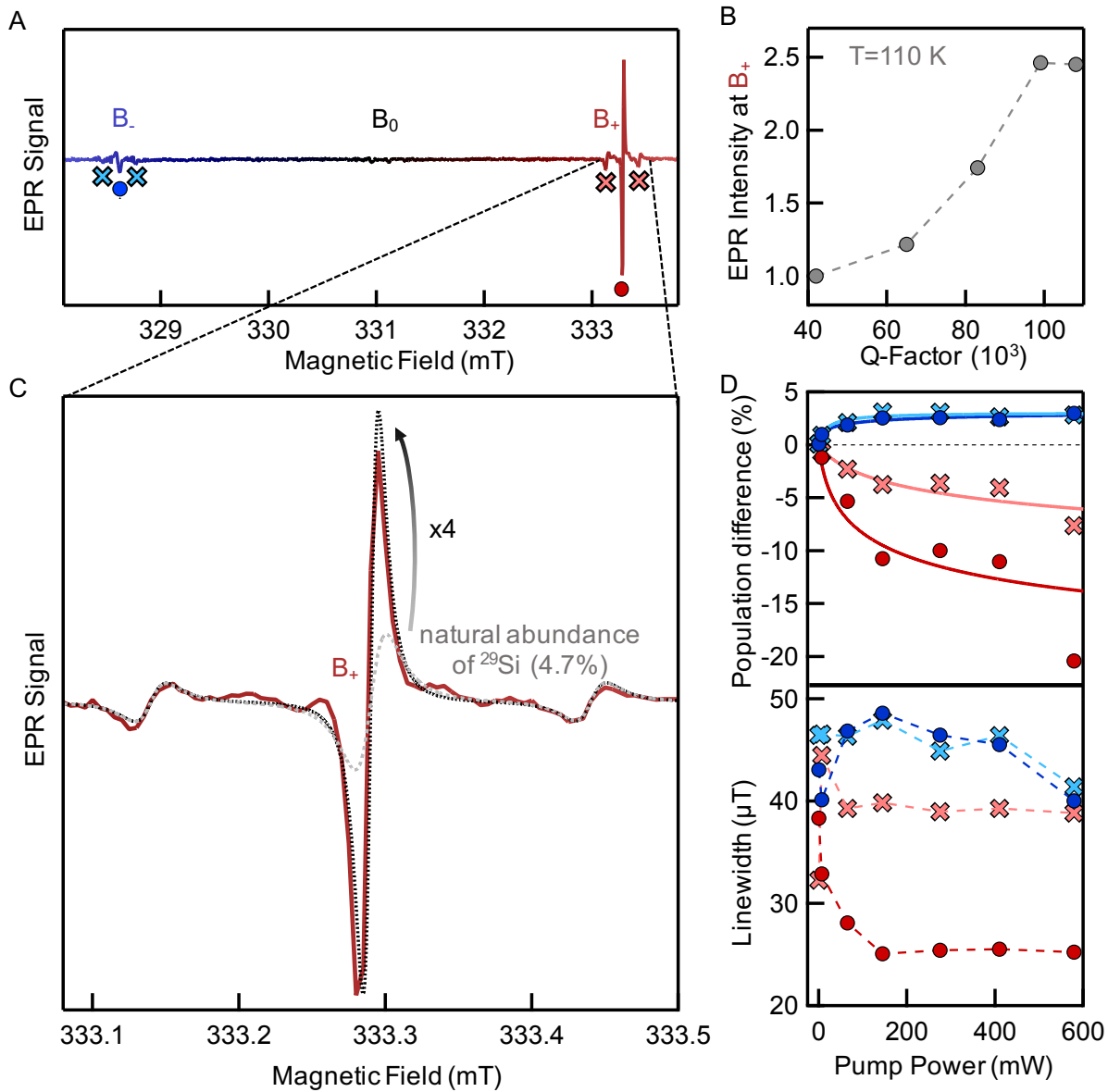

**Figure 4:** Stimulated emission and signs of superradiance for optimized settings: **(A)** EPR spectrum at $T = 110$ K for a high-Q resonator ($Q \approx 84000$). The emissive transition $B_+$ is much more intense compared to the absorptive transition $B_-$. **(B)** EPR intensity of the emissive transition $B_+$ for different Q-factors. **(C)** Zoomed-in region of the EPR spectrum to illustrate the HF structure of the emissive transition (experimental data in red with black dotted fit). The expected trace based on the natural abundance of $^{29}$Si is shown in grey. The experimentally observed intensity of the main peak deviates strongly from the expectations and is 4 times more intense than expected according to the isotope ratio of silicon nuclei. **(D)** Population difference and linewidth as a function of laser pump power. Circles represent the main peak of the $B_+$ (red) and $B_-$ (blue) EPR transitions, while crosses represent the values for the HF satellites. The $B_+$ linewidth collapses by a factor of 0.5 upon optical pumping and also the population difference of $B_+$ vastly exceeds that of $B_-$. The increased stimulated emission together with the linewidth collapse are signs of superradiance.

## Conclusion

In summary, on the way to the realization of a SiC-based maser, we have experimentally identified several physical parameters which have a strong impact on the spin polarization and, consequently, on the population inversion. They are related to the properties of the spin system itself as well as to the quality of the resonator hosting the crystal. Investigating the dependence of the spin polarisation on the excitation wavelength, we found that at low temperature ($T = 50$ K) the spin defect can be efficiently pumped via resonant excitation into the ZPL resulting in a higher population difference. At higher temperatures, i.e., when the ZPL has vanished, a Stokes excitation ($\lambda_{exc} > \lambda_{ZPL}$) is sufficient. Whether anti-Stokes excitation, where the sample temperature exceeds 300 K due to pump heating, would be more suitable for a potential SiC maser requires further investigation. Furthermore, we have analysed the influence of the spin defect orientation in the external magnetic field $\vec{B}$ on the population inversion. Here, we have found a linear relationship between the splitting of the absorptive and emissive EPR transitions and the population inversion. This alignment effect leads to an enhancement of the spin polarization by a factor of 2 at $\vec{B} \parallel \vec{c}$ compared to $\vec{B} \perp \vec{c}$ geometry, commonly used in EPR studies and should definitely be considered when designing a maser. Another fundamental requirement for the realization of a maser is a resonator with a high Q-factor and we demonstrate its enormous influence on the population inversion. Finally, we have shown that a variation in the number of contributing spins has a large effect on the population inversion and correspondingly a reduced linewidth resulting in the emergence of a superradiance of the emissive EPR transition. This could be done by carefully analysing the intensity ratio of HF peaks arising from interactions with spin-bearing and spin-less Si isotopes in a sample. With our results, we show how the spin polarization of a $S = 3/2$ spin system can be tuned to high values, and how it can be efficiently coupled to a high-Q microwave cavity. This altogether results in superradiance, foreshadowing the maser threshold in a technologically attractive SiC.

## Materials and Methods

### Silicon Carbide Samples

The investigated defects were created by 2 MeV electron irradiation ($3 - 10 \cdot 10^{17}$ cm$^{-2}$) of a 4H SiC wafer (Kasper et al. 2020). In this work we used two different samples: First, a small wafer piece (5.9 mm$^3$) with a defect density of $5.68 \cdot 10^{14}$ cm$^{-3}$. This sample is ideal to study the influence of excitation wavelength and defect orientation since it fits into standard EPR spectrometers (measurements shown in Figs. 2, 3). Furthermore, we investigated a stack of SiC wafers (34.3 mm$^3$) with a higher defect density of $2.27 \cdot 10^{15}$ cm$^{-3}$ for a larger ensemble of $\approx 7.8 \cdot 10^{13}$ V2 spin defects (Figs. 1, 4).

### Electron Paramagnetic Resonance

The EPR measurements were performed in three different setups with different advantages: The measurements for Figs. 1 and 4 were performed in a self-built EPR spectrometer as illustrated in Fig. 1B. Here, we were able to perform measurements on a large ensemble in a customized high-Q resonator. Measurements for Fig. 2B were performed in a modified Bruker spectrometer (E300) because the cavity (Bruker ER4104OR) possesses the best optical access for the tuneable laser (Sacher Lion 920). In order to perform angle-dependent measurements (see Fig. 3) we used a commercial benchtop spectrometer (Bruker Magnettech ESR5000) with a motorized precision goniometer.

Photoluminescence Spectroscopy

The PL measurements (Fig. 2A) were performed with a commercial LabRAM HR800 (Horiba) confocal micro Raman spectrometer with 633 nm laser excitation. The temperature was varied by a MicrostatHe (Oxford Instruments) cold finger cryostat.

## Conflict of Interest

The authors declare that the research was conducted in the absence of any commercial or financial relationships that could be construed as a potential conflict of interest.

## Author Contributions

AG and SS performed the wavelength and orientation dependent EPR studies. CK and AS determined optimum irradiation conditions. VB designed the microwave resonator. MK designed and realized the home-made spectrometer together with AG and AS. MW performed theoretical calculations of the relevant parameters affecting the maser threshold. MW and MK contributed equally to this work. AG, GA, VD, and AS wrote the manuscript. AS conceived the idea and supervised the project. All authors discussed the results and commented on the manuscript.

## Funding

This publication was supported by the Open Access Publication Fund of the University of Wuerzburg and the German Science Foundation (DFG) under the project DY 18/13-1.

## Data Availability Statement

The data that support the findings of this study are available on request from the corresponding author.